\documentclass[a4paper,English,manuscript]{revtex4}
\usepackage[latin9]{inputenc}
\usepackage{amsmath}
\usepackage{amssymb}
\usepackage{esint}
\usepackage{microtype}

\makeatletter

\@ifundefined{textcolor}{}
{%
 \definecolor{BLACK}{gray}{0}
 \definecolor{WHITE}{gray}{1}
 \definecolor{RED}{rgb}{1,0,0}
 \definecolor{GREEN}{rgb}{0,1,0}
 \definecolor{BLUE}{rgb}{0,0,1}
 \definecolor{CYAN}{cmyk}{1,0,0,0}
 \definecolor{MAGENTA}{cmyk}{0,1,0,0}
 \definecolor{YELLOW}{cmyk}{0,0,1,0}
}

\makeatother

\begin{document}
\title{Relativistic Kelvin circulation theorem for ideal Magnetohydrodynamics}
\author{Jianfei Wang}
\address{\textit{Department of Modern Physics, University of Science and Technology
of China, Hefei 230026, China}}
\author{Shi Pu}
\address{\textit{Department of Modern Physics, University of Science and Technology
of China, Hefei 230026, China}}
\begin{abstract}
We have studied the relativistic Kelvin circulation theorem for ideal
Magnetohydrodynamics. The relativistic Kelvin circulation theorem
is a conservation equation for the called $T$-vorticity, 
We have briefly reviewed the ideal magnetohydrodynamics in relativistic
heavy ion collisions. The highlight of this work is that we have obtained
the general expression of relativistic Kelvin circulation theorem
for ideal Magnetohydrodynamics. We have also applied the analytic
solutions of ideal magnetohydrodynamics in Bjorken flow to check our
results. Our main results can also be implemented to relativistic
magnetohydrodynamics in relativistic heavy ion collisions.
\end{abstract}
\maketitle

\section{Introduction}

Ideal magnetohydrodynamics (MHD) has been widely used in astrophysics
and plasma physics. Recently, it has also been applied to the relativistic
heavy ion collisions. In relativistic heavy ion collisions, two nuclei
collides with each other and generate an extreme strong electromagnetic
fields. The order of the magnetic fields can be $10^{17}-10^{18}$Gauss
\cite{Bzdak:2011yy,Deng:2012pc,Roy:2015coa,Li:2016tel}.

Many novel quantum transport effects have been introduced and studied
in such strong electromagnetic fields. The most important phenomenon
is named the chiral magnetic effect \cite{Vilenkin1980a,Kharzeev:2007jp,Fukushima:2008xe}.
It means that the magnetic fields can induce a charge current. The
microscopic theory for chiral magnetic effect is the quantum kinetic
theory for the massless fermions, named chiral kinetic theory \cite{Stephanov:2012ki,Son:2012wh,Gao:2012ix,Chen:2012ca,Gao:2015zka,Hidaka:2016yjf,Hidaka:2017auj,Gao:2017,Hidaka:2018mel}.
Besides, there are many other quantum transport phenomena. For example,
chiral electric separation effect means that the magnetic fields can
also induce the chiral current \cite{Huang:2013iia,Pu:2014cwa,Jiang:2014ura,Pu:2014fva}.
Very recently, it has been found that chirality production has a deep
connection to the famous Schwinger mechanism \cite{Fukushima:2010vw,Warringa:2012bq}.
We have used the world-line formalism to obtain the mass correction
to the axial Ward identity \cite{Copinger:2018ftr} and found that
the chirality production rate is the Schwinger pair production rate.
We have also discussed the other nonlinear effects coupled to the
electromagnetic fields \cite{Pu:2014cwa,Pu:2014fva,Chen:2016xtg,Hidaka:2018mel}.
One can find more the references therein.

On the other hand, there are also several studies on the analytic
solutions of MHD. Very recently, we have also obtained several analytic
solutions for the ideal MHD. In Ref. \cite{Pu:2016ayh,Roy:2015kma},
we have derived the solutions for the ideal MHD with transverse magnetic
fields in a Bjorken flow with and without magnetization effects. In
Ref. \cite{Pu:2016bxy,Pu:2016rdq}, we have extended the discussion
to the 2+1 dimensional Bjorken flow with transverse expansion. We
have also discussed the possible modification on $v_{2}$ induced
by the magnetic fields in Ref. \cite{Roy:2017yvg}. Later, we have
obtained the analytic solution for the anomalous MHD with transverse
magnetic fields in the presence of the chiral magnetic effects and
chiral anomaly \cite{Siddique:2019gqh}.

In relativistic heavy ion collisions, the called $T$-vorticity is
found to be conserved due to the relativistic Kelvin circulation theorem
\cite{PRL105,Gao:2014coa,Yang:2017asn}. In Ref. \cite{PRL105,Gao:2014coa},
authors have proved that in the ideal hydrodynamics without conserved
charged particles the the circular integration of $T$-vorticity is
proved to be conserved, which is called relativistic Kelvin circulation
theorem. Later, in Ref. \cite{Yang:2017asn}, we have extended the
discussion to the dissipative hydrodynamics.

In the work, we will extend the discussion on the relativistic Kelvin
circulation theorem to the ideal MHD.

The organization of the paper is as follows. In Sec. 2 
we will briefly review the ideal MHD. In Sec.3 
we have derived the general expression for the relativistic Kelvin
circulation theorem for the ideal MHD. We will conclude and discuss
in Sec.4 and Sec. 5.

Throughout this work, we will use the metric $g_{\mu\nu}=\mathrm{diag}\{+,-,-,-\}$,
thus, the fluid velocity satisfies $u^{\mu}u_{\mu}=1$, and the orthogonal
projector to the fluid four-velocity is $\Delta^{\mu\nu}=g^{\mu\nu}-u^{\mu}u^{\nu}$.
We also choose Levi-Civita tensor satisfying $\epsilon^{0123}=-\epsilon_{0123}=+1$.

\section{Ideal Relativistic Magnetohydrodynamics \label{subsec:Ideal-uncharged-fluid}}

In this section, we will review the ideal MHD. For more detials, one
can see Ref. \cite{Pu:2016ayh,Roy:2015kma,Pu:2016bxy,Pu:2016rdq,Siddique:2019gqh}.

In relativistic hydrodynamics, the electromagnetic strength tensor
$F^{\mu\nu}$ will be decomposed by the fluid velocity $u^{\mu}$,
\begin{equation}
F^{\mu\nu}=E^{\mu}u^{\nu}-E^{\nu}u^{\mu}+\epsilon^{\mu\nu\alpha\beta}u_{\alpha}B_{\beta},\label{decompose_EM}
\end{equation}
which gives, 
\begin{equation}
E^{\mu}=F^{\mu\nu}u_{\nu},\;B^{\mu}=\frac{1}{2}\epsilon^{\mu\nu\alpha\beta}F_{\alpha\beta}.
\end{equation}
Generally, we can parametrize the velocity as, $u^{\mu}=\gamma(1,\mathbf{v}),$with
$\gamma$ being $1/\sqrt{1-|\mathbf{v}|^{2}}$. We can prove that,
\begin{equation}
u\cdot E=u\cdot B=0.
\end{equation}
In the comoving frame of a fluid cell, i.e. local rest frame, $u^{\mu}=(1,\mathbf{0})$,
the $E^{\mu}$ and $B^{\mu}$ will reduce to 
\begin{equation}
E^{\mu}=(0,\mathbf{E}),B^{\mu}=(0,\mathbf{B}).
\end{equation}

The ideal MHD includes the energy-momentum and charge conservation
equations coupled to the Maxwell's equations.

The energy-momentum conservation is given by, 
\begin{equation}
\partial_{\mu}T^{\mu\nu}=0,\label{eq:consevation_01}
\end{equation}
where $T^{\mu\nu}$ is the energy-momentum tensor. In an ideal fluid,
this tensor can be decomposed as two parts, 
\begin{equation}
T^{\mu\nu}=T_{matter}^{\mu\nu}+T_{EM}^{\mu\nu}.
\end{equation}
The first term is the part for the matter 
\begin{equation}
T_{matter}^{\mu\nu}=(\epsilon+P)u^{\mu}u^{\nu}-(P+\Pi)g^{\mu\nu}+\pi^{\mu\nu},
\end{equation}
where $\epsilon,P$ are energy density and pressure. $\Pi$ and $\pi^{\mu\nu}$
denote the bulk pressure and shear viscous tensor, respectively. In
the first order relativistic hydrodynamics, $\Pi$ and $\pi^{\mu\nu}$
can be parameterized as, 
\begin{eqnarray}
\Pi & = & \zeta(\partial\cdot u),\\
\pi^{\mu\nu} & = & 2\eta[\frac{1}{2}\Delta^{\mu\alpha}\Delta^{\nu\beta}+\frac{1}{2}\Delta^{\nu\alpha}\Delta^{\mu\beta}\nonumber \\
 &  & -\frac{1}{3}\Delta^{\mu\nu}\Delta^{\alpha\beta}]\partial_{\alpha}u_{\beta},
\end{eqnarray}
where $\zeta$ and $\eta$ stand for the bulk and shear viscosities,
respectively. The projector is defined as, 
\begin{equation}
\Delta^{\mu\nu}=g^{\mu\nu}-u^{\mu}u^{\nu}.
\end{equation}
It is easy to prove that 
\begin{eqnarray}
\Delta^{\mu\nu}u_{\nu} & = & \Delta^{\mu\nu}u_{\mu}=0,\nonumber \\
\Delta^{\mu\alpha}\Delta_{\alpha}^{\nu} & = & \Delta^{\mu\nu}.
\end{eqnarray}
where we have used that $u^{\mu}u_{\nu}=1$.

Both $\Pi$ and $\pi^{\mu\nu}$ are dissipative effects. In an ideal
fluid, all the dissipative terms vanish, i.e. $\Pi=0$ and $\pi^{\mu\nu}=0$.

The second part for the energy-momentum tensor comes from the electromagnetic
fields, 
\begin{eqnarray}
T_{EM}^{\mu\nu} & = & -F^{\mu\lambda}F_{\lambda}^{\nu}+\frac{1}{4}g^{\mu\nu}F^{\rho\sigma}F_{\rho\sigma}\nonumber \\
 & = & -E^{\mu}E^{\nu}-B^{\mu}B^{\nu}+(E^{2}+B^{2})u^{\mu}u^{\nu}\nonumber \\
 &  & -\frac{1}{2}g^{\mu\nu}(E^{2}+B^{2})+u^{\mu}\epsilon^{\nu\lambda\rho\sigma}E_{\lambda}u_{\rho}B_{\sigma}\nonumber \\
 &  & +u^{\nu}\epsilon^{\mu\lambda\rho\sigma}E_{\lambda}u_{\rho}B_{\sigma},
\end{eqnarray}
where 
\begin{eqnarray}
E^{2} & \equiv-E^{\mu}E_{\mu},\nonumber \\
B^{2} & \equiv-B^{\mu}B_{\mu}.
\end{eqnarray}

Inserting Eq.(\ref{decompose_EM}) into the above equation yields,
\begin{eqnarray}
T^{\mu\nu}= & (\epsilon+p+E^{2}+B^{2})u^{\mu}u^{\nu}\nonumber \\
 & -(p+\frac{1}{2}E^{2}+\frac{1}{2}B^{2})g^{\mu\nu}\nonumber \\
 & -E^{\mu}E^{\nu}-B^{\mu}B^{\nu}\nonumber \\
 & +u^{\mu}\epsilon^{\nu\lambda\rho\sigma}E_{\lambda}u_{\rho}B_{\sigma}\nonumber \\
 & +u^{\nu}\epsilon^{\mu\lambda\rho\sigma}E_{\lambda}u_{\rho}B_{\sigma}.\label{EMT_01}
\end{eqnarray}

The charge conservation equation reads, 
\begin{equation}
\partial_{\mu}j^{\mu}=0.
\end{equation}
The charge current $j^{\mu}$ can be decomposed as, 
\begin{equation}
j^{\mu}=nu^{\mu}+\sigma E^{\mu},\label{eq:current_01}
\end{equation}
where $n$ is the electric charge density and $\sigma$ is the electric
conductivity.

For simplicity, we consider a charge neutral fluid, i.e. 
\begin{equation}
n=0.
\end{equation}

In order to simplify the calculations, we will use the ideal MHD limit.
The ideal MHD limit means that the electric conductivity is infinite,
i.e. 
\begin{equation}
\sigma\rightarrow\infty.
\end{equation}
To avoid the divergence in $\sigma E^{\mu}$ term in Eq. (\ref{eq:current_01}),
the four vector form of $E^{\mu}$ must vanish, 
\begin{equation}
E^{\mu}=0.
\end{equation}

In the ideal limit, i.e. $E^{\mu}=0$, the energy-momentum tensor
in Eq.(\ref{EMT_01}) reduces to, 
\begin{eqnarray}
T^{\mu\nu} & =(\epsilon+P+B^{2})u^{\mu}u^{\nu}-(P+\frac{1}{2}B^{2})g^{\mu\nu}-B^{\mu}B^{\nu}\nonumber \\
\label{eq:EMT_01}
\end{eqnarray}

Now, we will discuss the covariant form of Maxwell's equation, 
\begin{align}
\partial_{\mu}(\epsilon^{\mu\nu\alpha\beta}F_{\alpha\beta}) & =0,\nonumber \\
\partial_{\mu}F^{\mu\nu} & =j^{\nu}.
\end{align}

In ideal MHD, the covariant equations for magnetic fields are given
by, 
\begin{equation}
\partial_{\nu}(B^{\mu}u^{\nu}-B^{\nu}u^{\mu})=0.\label{eq:Maxwell_02}
\end{equation}
In the local rest frame, i.e. $u^{\mu}=(1,\mathbf{0})$, the above
equation reduces to 
\begin{equation}
\frac{d}{dt}\mathbf{B}+\mathbf{B}(\nabla\cdot\mathbf{v})-(\mathbf{B}\cdot\nabla)\mbox{\textbf{v}}=0.
\end{equation}
Contracting Eq. (\ref{eq:Maxwell_02}) with $u_{\mu}$, yields, 
\begin{equation}
\partial_{\mu}B^{\mu}=-B^{\mu}(u\cdot\partial)u_{\mu},
\end{equation}
which gives 
\begin{equation}
\nabla\cdot\mathbf{B}=0,
\end{equation}
in the local rest frame. Contracting Eq. (\ref{eq:Maxwell_02}) with
$B_{\mu}$, yields, 
\begin{equation}
\frac{1}{2}(u\cdot\partial)B^{2}+B^{2}(\partial\cdot u)+B^{\mu}B^{\nu}\partial_{\nu}u_{\mu}=0.\label{eq:Maxwell_03}
\end{equation}

The thermodynamical relations reads, 
\begin{equation}
\epsilon+p=Ts+\mu n,
\end{equation}
where $T,s,\mu,n$ are temperature, entropy density, chemical potential
and charge number density, respectively. In our case, since we have
considered the charge neutral fluid $n=0$, the thermodynamical relation
becomes, 
\begin{align}
\epsilon+p & =Ts.\label{eq:thermo_relation_01}
\end{align}
Its differential form is given by, 
\begin{align}
d\epsilon & =Tds,\nonumber \\
dp & =sdT.\label{eq:thermo_relation_02}
\end{align}

\section{Relativistic Kelvin circulation theorem \label{sec:Relativistic-Kelvin-circulation}}

In the section, we will derive the general expression for the relativistic
Kelvin circulation theorem for ideal MHD.

The projection of Eq. (\ref{eq:consevation_01}) along the velocity
$u^{\nu}$ reads, 
\begin{equation}
u_{\nu}\partial_{\mu}T^{\mu\nu}=0.
\end{equation}
Inserting Eq. (\ref{eq:EMT_01}) into above equation yields, 
\begin{align}
 & (u\cdot\partial)\epsilon+(\epsilon+p)(\partial\cdot u)\nonumber \\
 & +(u\cdot\partial)\frac{1}{2}B^{2}+B^{2}(\partial\cdot u)\nonumber \\
 & -u_{\mu}(B\cdot\partial)B^{\mu}=0,
\end{align}
where we have used, $u\cdot B=0$. Applying Eq. (\ref{eq:Maxwell_03}),
we eventually obtain, 
\begin{eqnarray}
(u\cdot\partial)\epsilon+(\epsilon+p)(\partial\cdot u) & = & 0.\label{eq:EM_conser_03}
\end{eqnarray}
Using the thermodynamical relations (\ref{eq:thermo_relation_01},
\ref{eq:thermo_relation_02}), we get, 
\begin{equation}
\partial_{\mu}(su^{\mu})=0.\label{eq:entropy_02}
\end{equation}
We can take a volume integral over Eq.(\ref{eq:entropy_02}) and obtain
that 
\begin{equation}
\frac{dS}{dt}=\int_{V}d^{3}x\partial_{\mu}(su^{\mu})=0,
\end{equation}
where $S$ is the total entropy for the whole system. The above equation
mean the total entropy is conserved.

Another equation for the energy-momentum conservation is 
\begin{equation}
\Delta_{\alpha\nu}\partial_{\mu}T^{\mu\nu}=0,
\end{equation}
which gives us, 
\begin{eqnarray}
(u\cdot\partial)u_{\alpha} & = & \frac{1}{(\epsilon+p+B^{2})}[\Delta_{\alpha}^{\nu}\partial_{\nu}(p+\frac{1}{2}B^{2})\nonumber \\
 & + & \Delta_{\mu\alpha}(B\cdot\partial)B^{\mu}+B_{\alpha}(\partial\cdot B)].
\end{eqnarray}
It tells us that the magnetic field may accelerate the fluid velocity.
However, in our previous work \cite{Pu:2016ayh,Roy:2015kma,Pu:2016bxy,Pu:2016rdq,Siddique:2019gqh},
we have found a kind of force-free type configuration for the magnetic
fields. In such cases, the fluid velocity will not be acceptilated.

We can also rewrite the $T^{\mu\nu}$ as, 
\begin{equation}
T^{\mu\nu}=(Ts+B^{2})u^{\mu}u^{\nu}-(P+\frac{1}{2}B^{2})g^{\mu\nu}-B^{\mu}B^{\nu}.
\end{equation}
Eq. (\ref{eq:consevation_01}) reads, with the help of Eq. (\ref{eq:thermo_relation_01}),
\begin{eqnarray}
 &  & su^{\mu}\partial_{\mu}(Tu^{\nu})-\partial^{\nu}(P+\frac{1}{2}B^{2})\nonumber \\
 & + & \partial_{\mu}(B^{2}u^{\mu}u^{\nu})-\partial_{\mu}(B^{\mu}B^{\nu})=0.
\end{eqnarray}
Using the thermodynamic relation (\ref{eq:thermo_relation_01}, \ref{eq:thermo_relation_02})
again, we get, 
\begin{align}
 & u^{\mu}\partial_{\mu}(Tu^{\nu})-\partial^{\nu}T\nonumber \\
 & =\frac{1}{s}\left[\partial^{\nu}(\frac{1}{2}B^{2})-\partial_{\mu}(B^{2}u^{\mu}u^{\nu})+\partial_{\mu}(B^{\mu}B^{\nu})\right].\label{eq:temp_cons_01}
\end{align}

For convenience, we can introduce an antisymmetric $T$-vorticity
tensor $\Xi^{\mu\nu}$ by \cite{PRL105,Gao:2014coa,Yang:2017asn},
\begin{equation}
\Xi^{\mu\nu}=\partial^{\nu}(Tu^{\mu})-\partial^{\mu}(Tu^{\nu}),\label{eq:thermal_vor_01}
\end{equation}
which is like $F^{\mu\nu}$ in an electromagnetic field. In this case,
we can use a compact form to rewrite Eq. (\ref{eq:temp_cons_01}),
\begin{align}
 & \Xi^{\mu\nu}u_{\nu}\nonumber \\
 & =\frac{1}{s}\left[\partial^{\mu}(\frac{1}{2}B^{2})-\partial_{\nu}(B^{2}u^{\mu}u^{\nu})+\partial_{\nu}(B^{2}b^{\mu}b^{\nu})\right].\label{eq:conservation_eq_01}
\end{align}

Now, we will discuss the integral form of the $T$-vorticity. By introducing
the proper time $\tau$ with the relation $d/d\tau=u^{\mu}\partial_{\mu}$,
the above equation can be expressed as a circulation integral along
a covariant loop $L(\tau)$, 
\begin{eqnarray}
 &  & \frac{d}{d\tau}\oint_{L(\tau)}Tu^{\mu}dx_{\mu}\nonumber \\
 & = & \oint_{L(\tau)}\frac{d}{d\tau}(Tu^{\mu})dx_{\mu}\nonumber \\
 & = & \oint_{L(\tau)}u^{\nu}\partial_{\nu}(Tu^{\mu})dx_{\mu}\nonumber \\
 & = & \oint_{L(\tau)}\Xi^{\mu\nu}u_{\nu}dx_{\mu}\nonumber \\
 & = & \oint_{L(\tau)}\frac{1}{s}\left[\partial^{\nu}(\frac{1}{2}B^{2})-\partial_{\mu}(B^{2}u^{\mu}u^{\nu})\right..\nonumber \\
 & + & \left.\partial_{\mu}(B^{2}b^{\mu}b^{\nu})\right]dx_{\nu}\label{eq:Kelvin_MHD_01}
\end{eqnarray}
In the first line, the differential acts on the $dx^{\mu}$ is nonzero,
i.e. $\frac{d}{d\tau}dx^{\mu}=d\left(\frac{dx^{\mu}}{d\tau}\right)=du^{\mu}=dx^{\nu}\partial_{\nu}u^{\mu}$,
but it vanishes since $u^{\mu}\partial_{\nu}u_{\mu}dx^{\nu}=0$. In
the second line we have used the Stokes theorem, 
\begin{equation}
\oint_{L(\tau)}\partial^{\mu}Tdx_{\mu}=\oint_{S}[\partial^{\mu}\partial^{\nu}T-\partial^{\nu}\partial^{\mu}T]d\sigma_{\mu\nu}=0.
\end{equation}
Eq. (\ref{eq:Kelvin_MHD_01}) is the relativistic Kelvin circulation
theorem for ideal MHD. It means that the magnetic fields may modify
the $T$-vorticity consideration.

\section{Discussion \label{sec:Discussion-and-conclusion}}

We will apply the main result in Eq.(\ref{eq:Kelvin_MHD_01}) to the
one of the analytic solution found in Ref. \cite{Pu:2016ayh,Roy:2015kma}.

In our work \cite{Pu:2016ayh,Roy:2015kma}, we consider the ideal
MHD with transverse magnetic fields in a Bjorken flow. The Bjorken
flow means that the fluid velocity is given by, 
\begin{equation}
u^{\mu}=(\frac{t}{\tau},0,0,\frac{z}{\tau}),
\end{equation}
where $\tau$ is the proper time defined as, 
\begin{equation}
\tau=\sqrt{t^{2}-z^{2}}.
\end{equation}
All the thermodynamic quantities, such as energy density, pressure,
temperature, entropy density etc., will only depend on the proper
time. The magnetic fields decays as, 
\begin{equation}
B^{\mu}=(0,0,B_{0}\frac{\tau_{0}}{\tau},0),
\end{equation}
where $B_{0}$ and $\tau_{0}$ are the initial magnetic field and
initial proper time, respectively.

Inserting this solution into Eq. (\ref{eq:Kelvin_MHD_01}), we find
that the 
\begin{equation}
\frac{d}{d\tau}\oint_{L(\tau)}Tu^{\mu}dx_{\mu}=0.
\end{equation}
for this special solutions. In this case, the $T$-vorticity is still
conserved.

\section{Conclusion}

\label{conclusion}

In this paper, we have derived the general expression for the relativistic
Kelvin circulation theorem for ideal MHD. Firstly, we have briefly
reviewed the conservation equations and Maxwell's equations in the
ideal MHD. We introduce the ideal limit for MHD, i.e. the electric
conductivity is infinite. In this limit, the $E^{\mu}$ vanishes to
ensure the charge current finite. Then, we have rewritten the energy-momentum
conservation equation by using the thermodynamic relations. We have
also introduce the $T$-vorticity, $\Xi^{\mu\nu}$ in Eq.(\ref{eq:thermal_vor_01}).
Next, we have derived the relativistic Kelvin circulation theorem
for ideal MHD shown in Eq. (\ref{eq:Kelvin_MHD_01}). At last, We
have also checked the results for the ideal MHD in a Bjorken flow.

In the future, we will extend the discussion to the normal MHD and
anomalous MHD. We can also consider the magnetization effects in this
framework.

\end{document}